# Strong superconducting pairing strength and pseudogap features in a putative multiphase heavy-fermion superconductor CeRh$_2$As$_2$ by soft point-contact spectroscopy

Qingxin Dong[1,2=], Tong Shi[1,3=], Pengtao Yang[1,2=], Xinyang Liu[1,4], Xiaofan Shi[1,2], Lei Wang[1,2], Junsen Xiang[1,2], Hanming Ma[1,2], Zhaoming Tian[3], Jianping Sun[1,2], Yoshiya Uwatoko[5], Genfu Chen[1,2,6], Xinbo Wang[1,2], Jie Shen[1,2], Rui Wu[1], Xin Lu[7], Peijie Sun[1,2], Grzegorz Chajewski[8], Dariusz Kaczorowski[8*], Bosen Wang[1,2*] and Jinguang Cheng[1,2*]

[1]*Beijing National Laboratory for Condensed Matter Physics and Institute of Physics, Chinese Academy of Sciences, Beijing 100190, China*

[2]*School of Physical Sciences, University of Chinese Academy of Sciences, Beijing 100049, China*

[3]*Wuhan National High Magnetic Field Center and School of Physics, Huazhong University of Science and Technology, Wuhan 430074, China*

[4]*School of physics, Beihang University, Beijing 100191, China*

[5]*Institute for Solid State Physics, University of Tokyo, Kashiwanoha 5-1-5, Kashiwa, Chiba 277-8581, Japan*

[6]*Songshan Lake Materials Laboratory, Dongguan, Guangdong 523808, China*

[7]*Center for Correlated Matter, School of Physics, Zhejiang University, Hangzhou 310058, China*

[8]*Institute of Low Temperature and Structure Research, Polish Academy of Sciences, Okólna 2, 50-422 Wrocław, Poland*

= These authors contributed equally to this work.

*Corresponding authors: bswang@iphy.ac.cn (BSW);

d.kaczorowski@intibs.pl (DK);  jgcheng@iphy.ac.cn (JGC)

## Abstract

CeRh$_2$As$_2$ is a newly discovered candidate of multiphase heavy-fermion superconductor ($T_c \approx 0.3$ K) with intriguing physical properties. Here, we employ soft point-contact spectroscopy to investigate its energy gap behaviors in both the normal and superconducting states. The differential conductance below $T_c$ reveals an estimated superconducting energy gap of $2\Delta_{SC} \approx 0.24$ meV and thus an extremely strong superconducting pairing strength $2\Delta_{SC}/k_BT_c \approx 8.8$, which is comparable to those of cuprates and iron-based high-$T_c$ superconductors as well as infinite-layer nickelates. Above $T_c$, a well-defined pseudogap feature is manifested as a $V$-shaped dip in the differential conductance spanning an energy scale of $2\Delta_g \approx 0.95$-$3.0$ meV. The pseudogap feature persists to the highest characteristic temperature of $T_g \approx 8$-$9$ K and is gradually suppressed by magnetic field of $B_g \approx 9.0 \pm 0.5$ T regardless of its direction relative to the crystallographic axes. The observation of pseudogap features prior to the superconducting phase transition enriches the phase diagram of CeRh$_2$As$_2$ and provides a novel platform to study the interplay of unconventional superconductivity and pseudogap phenomena.

Recently, CeRh$_2$As$_2$, a heavy-fermion superconductor (HFSC), has attracted much attention because its superconducting state below $T_c \approx 0.3$ K was found to undergo a magnetic-field-induced phase transition[1-3]. This intriguing observation has been attributed to a transition from low-field even-parity SC1 to high-field odd-parity SC2 phase [2-10]. This makes CeRh$_2$As$_2$ another example among HFSCs possessing multiple superconducting phases besides UTe$_2$[11,12] and UPt$_3$[13,14]. Interestingly, the SC2 phase only exists when the magnetic field is applied parallel to the $c$-axis ($B//c$). The upper critical field $B_{c2}(0)$ is anisotropic and as large as $B_{c2}(0) \approx 14$ T for $B//c$ [1,7], far beyond the Pauli limit $B_p = 1.84T_c$. It has been proposed that these distinctive physical properties of CeRh$_2$As$_2$ are rooted in its unique crystal structure, characterized by a globally centrosymmetric unit cell in which the inversion symmetry is locally broken in the sublayers [11,12]. As shown in Fig. S1(a) of the Supplementary Materials (SM), CeRh$_2$As$_2$ adopts a tetragonal CaBe$_2$Ge$_2$-type structure with alternatively stacked Ce-atom layers and two different Rh-As layer blocks along the $c$ axis. It was suggested that the staggered Rashba spin-orbit coupling between alternating Ce sub-layers introduces sub-lattice degrees of freedom and serves as the origin of field-induced superconducting transition at about 4.0 T and the observed large $B_{c2}(0)$ of SC2 phase [11,12].

Motivated by the above discovery, follow-up studies have uncovered more exotic properties in the normal and superconducting states of CeRh$_2$As$_2$. For example, a possible quadrupole density wave (QDW) order at $T_0 \approx 0.40$ K was proposed based on the measurements of thermal expansion, specific heat and x-ray spectroscopy [5,15]. However, a recent study[8] on the super-higher-quality samples tends to attribute the features at $T_0$ to a long-range antiferromagnetic (AFM) order in analogy to other HFSCs, such as CeRhIn$_5$ and Ce$_2$RhIn$_8$[16,17]. It revealed a first-order transition within the superconducting state, which was attributed to a metamagnetic transition, in accordance with the earlier nuclear quadrupole resonance (NQR) and nuclear magnetic resonance (NMR) measurements demonstrating the existence of the AFM order below $T_N \approx 0.25$ K[2,6]. Subsequent muon spin relaxation ($\mu$SR) experiments confirmed the coexistence of quasi-local AFM magnetism with bulk SC[4]. In turn, angle-resolved photoemission spectroscopy (ARPES) revealed the presence of Fermi surface nesting[10] and suggested a potential link to magnetic excitations or QDW phenomena. In contrast to the initially suggested scenario of a spin-triplet nature of the high-field odd-parity SC2 state, the studies on superconducting spin susceptibility at two nonequivalent As sites have argued a spin-singlet state for both SC1 and SC2 phases[2,6]. They pointed out that the AFM order only coexists within the low-field-SC1 phase with no sign of magnetic ordering in the SC2 phase[2,6]. These controversial results call for further studies on the superconducting and normal states of CeRh$_2$As$_2$, especially at the microscopic level using spectroscopic probes.

In this Letter, we report on our detailed study on the microscopic properties of high-quality CeRh$_2$As$_2$ single crystals in both normal and superconducting states using soft point-contact spectroscopy (PCS) under extreme conditions, at temperatures down to 50 mK and high magnetic fields up to 12 T. Our results revealed the superconducting energy gap of $2\Delta_{SC} \approx 0.24$ meV at 50 mK and strong superconducting pairing strength $2\Delta_{SC}/k_B T_c \approx 8.8$, which is comparable to those of the copper/iron-based high-$T_c$ SCs. Moreover, above $T_c$, we found a well-defined pseudogap feature, manifested as a V-shaped dip in PCS spectra with the energy scale of $2\Delta_g \approx 0.95$-3.0 meV, persisting up to high temperature of $T_g \approx 8$-9 K and large critical field of $B_g \approx 9.0 \pm 0.5$ T applied along the $c$ or $a$ crystallographic axes. The discovery of pseudogap feature in CeRh$_2$As$_2$, reminiscent of HFSC CeCoIn$_5$[18] and high-$T_c$ cuprates[19], provides new clues for elucidating the underlying mechanisms of unconventional SC in this material.

The CeRh$_2$As$_2$ single crystals used in this study were grown by a flux method as described previously[7,8]. The results of detailed characterizations of their crystal-chemical quality are summarized in Note 1 of SM. The X-ray diffraction (XRD) and Laue backscattering patterns confirmed the tetragonal CaBe$_2$Ge$_2$-type structure (space group P4/*nmm*), Fig. S1(b)-(d). The chemical composition determined by an energy dispersive X-ray spectroscopy (EDS) yielded the ratio 18.90 : 40.48 : 40.62 of the Ce, Rh and As constituents, which is close to the ideal 1:2:2 one. The EDS mapping revealed a homogeneous distribution of the elements, Fig. S1(e). The electrical resistivity $\rho(T)$ measured on three different single-crystalline specimens (S1#, S2#, S3#) reproducibly displayed a sharp superconducting transition at $T_c \approx 0.3$ K. The estimated residual resistivity ratio $\rho(T_m)/\rho(0.5K)$ was similar to the reported values [1,3,5,8], Fig. S2. All data confirmed a superior quality of the $CeRh_2As_2$ crystals used in our PCS studies.

Soft PCS was performed by attaching 20 $\mu$m-diameter gold wires with a silver-paint drop on the (00*l*) crystalline surfaces. In such a configuration, thousands of parallel nanoscale channels form between individual silver particles and the surface. The differential conductance (d$I$/d$V$) spectra were collected as a function of bias voltage with a lock-in technique in a quasi-four-probe configuration, Fig.1(a). The output current was mixed with both dc and ac components supplied by a 6221 Keithley current source and a SR830 lock-in amplifier, respectively. The first harmonic response of the lock-in amplifier is proportional to the bias voltage dependent point-contact resistance. All the experiments were performed using either a $^4$He cryostat or a dilution refrigerator equipped with a superconducting magnet, available at the Synergic Extreme Condition User Facility (SECUF), Beijing.

Fig.1(b) presents the d$I$/d$V$ spectra of $CeRh_2As_2$ (S1#), taken in the temperature interval 50-300 mK with measuring current $I_{AC}$ = 50 $\mu$A. The d$I$/d$V$ spectrum exhibits a bimodal feature below 200 mK, above which, it gradually evolves into a single peak and weakens in intensity until vanishes completely above 300 mK. This feature signals opening of the superconducting gap below 300 mK, in concert with the superconducting transition in $CeRh_2As_2$ (SC1 phase) determined from the $C_p(T)/T$, $\rho(T)$ and $\chi(T)$[1,3,8,20-22]. Remarkably, the peak in d$I$/d$V$ is superimposed on a distinct *V*-shaped signal, which becomes increasingly noticeable especially above 280 mK. This extra feature in the d$I$/d$V$ spectra can be attributed to the occurrence of pseudogap as discussed later.

To identify the evolution of the superconducting gap 2$\Delta_{SC}$, the normalized (d$I$/d$V$)/(d$I$/d$V$)$_{0.3K}$ spectra were analyzed, Fig. 1(c). At small voltage bias, they exhibit a broad peak that disappears near 300 mK (to verify the intrinsic nature of this feature we measured the d$I$/d$V$ spectra for a few other junctions and found it sample independent, Fig. S3). As shown by the dashed lines in Fig. 1(c) and Fig. S3(b), the gap 2$\Delta_{SC}$ was found to monotonically decrease with increasing temperature, Fig. S3(c). Its magnitude at 50 mK is about 0.24 meV, which yields a very large superconducting pairing strength $2\Delta_{SC}/k_BT_c \approx 8.8$, comparable to the values reported for cuprates (~ 11 for $YBa_2Cu_3O_{6+\delta}$[23] and $Bi_2Sr_2CaCu_2O_{8+\delta}$[24]), iron-based high-$T_c$ superconductors (~ 8.12 for $(Li_{0.84}Fe_{0.16})OHFe_{0.98}Se$[25], ~ 6.6 for $Ba_{0.6}K_{0.4}Fe_2As_2$[26]) and infinite-layer nickelate systems (~ 5.8 for $Nd_{1-x}Sr_xNiO_2$ thin film[27]).

Fig. 1(d) shows the d$I$/d$V$ spectra of $CeRh_2As_2$ (S1#) taken at $T$ = 50 mK and different external magnetic fields $B$//$c$ with $I_{AC}$ = 50 μA. Clearly, the bimodal peak persists up to as large field as 5 T, transforms into a single peak feature near 6 T, and disappears above 7 T due to magnetic-field-driven Cooper pair breaking. Accordingly, the gap $2\Delta_{SC}$ gradually diminishes and drops to zero above 7 T, Fig. S3(d). The field dependence of $T_c$ was determined from the evolution of zero-bias voltage conductance peak (ZBCP), Fig.1(e). The so-defined superconducting $T_c^{ZBCP}$ gradually decreases with increasing field and disappears above 7 T.

Figs.1(f)-(g) present the PCS results of $CeRh_2As_2$ (S1#) measured at 50 mK with $I_{AC}$= 50 μA in magnetic field $B\perp c$. As can be inferred from these figures, the d$I$/d$V$ spectra and temperature-dependent ZBCP($T$) are similar to those obtained with $B$//$c$, except for quicker closing of $2\Delta_{SC}$ and faster decrease of $T_c^{ZBCP}$ (see a comparison in Fig. 3S(d)). The upper critical field estimated for $B\perp c$ from the $T_c^{ZBCP}$ versus $B$ equals $B_{c2}$ $\approx 2.0$ T at 50 mK, in good agreement with the literature [1,3,8], which demonstrates the reliability of PCS in characterizing the superconducting properties of $CeRh_2As_2$.

Figs. 4S(a)-(b) visualize the d$I$/d$V$ spectra measured at 50 mK with different $I_{AC}$ ranging from 50 μA to 500 μA. It is noteworthy that the superconducting signal significantly weakens with increasing $I_{AC}$, but the $V$-shaped pseudogap behavior remains unchanged, indicating different energy scales of these two features. In order to determine the temperature and magnetic field dependencies of the pseudogap, measurements of the d$I$/d$V$ spectra of $CeRh_2As_2$ (S2#) were performed using $I_{AC}$ = 100 μA that was selected to highlight the pseudogap characteristics. Fig. 2(a) presents the results obtained in zero magnetic field in the temperature range 60 mK-12 K. A well-defined $V$-shaped dip emerging at low-bias voltages, combined below $T_c$ with weak superconducting signal, indicates an extra energy gap in $CeRh_2As_2$. This feature can be attributed to formation of the pseudogap near Fermi level, similar to that occurring in high-$T_c$ cuprates[19]. With increasing temperature, the weak superconducting signal gradually disappears, while the $V$-shaped dip persists in the normal state up to as high temperature as $T_g \approx 8$-9 K. To test sample dependence of the pseudogap feature, we performed PCS experiments on three different samples (S1#, S2#, S3#) of $CeRh_2As_2$, Fig. S5. The results are similar and only subtle differences were noticed in the spectral shape, bias voltage, dip amplitude and background signal. On this basis one can firmly conclude that the pseudogap is an intrinsic property of $CeRh_2As_2$.

Generally, the asymmetry in the d$I$/d$V$ spectra observed at high-bias voltages can be attributed to the Fano effect[28-30]. To quantify the pseudogap behavior in CeRh$_2$As$_2$, we first deducted the asymmetric part of the d$I$/d$V$ spectra, as described in Note 6 of SM, getting good description of the PCS spectra at high-bias voltages and at temperatures above 12 K (see solid lines in Fig. 2(a)). The energy width of Fano resonance, estimated from a linear extrapolation, is $\gamma_0$ ($T$ = 0) ≈ 5.4 meV, Fig. S6. The so-obtained Kondo temperature $T_K$ = $\gamma_0$/$k_B$ ≈ 60 K is fairly close to 44 K determined from $\rho(T)$[1-3].

The normalized d$I$/d$V$ spectra after subtracting the Fano contributions are shown in Fig. 2(b). The pseudogap is characterized by two parameters: its width 2$\Delta_g$, determined by the bias voltage at half of the $V$-shaped peak, and its amplitude $A_g$, defined as the dip depth at zero-bias voltage. Figs. 2(c) and 2(d) display the field dependencies of the pseudogap feature measured on CeRh$_2$As$_2$ (S2#) in magnetic field $B$//$c$ with $I_{AC}$ = 100 μA at $T$ = 0.1 K (below $T_c$) and $T$ = 0.5 K (above $T_c$), respectively. At 0.1 K, the Andreev reflection peak occurring at the zero bias voltage becomes suppressed in fields stronger than 1 T, and the pseudogap has an energy scale of 2$\Delta_g$ ≈ 0.95 meV. With further increasing field, the uniform $V$-shaped feature weakens and vanishes near 9 T, i.e. in a field somewhat smaller than $B_{c2}$ ≈ 12 T (at 0.1 K) of the high-field SC2 phase. The pseudogap in the normal state is also gradually suppressed by magnetic field and completely disappears on approaching 9 T. As shown in Fig. 2(e), similar results were obtained at $T$ = 0.5 K for CeRh$_2$As$_2$ (S1#) at $B \perp c$.

As can be inferred from the figure, the normalized d$I$/d$V$ spectra ($I_{AC}$ = 100 μA) slightly differ from those obtained with $B$//$c$ only in terms of the values of critical field and energy scale. In order to verify the reproducibility of the pseudogap characteristics, the d$I$/d$V$ spectra were measured for several different junctions of three CeRh$_2$As$_2$ samples (S1#, S2#, S3#), Figs. S7-S11 ($B$//$c$) and Figs. S12-S13 ($B \perp c$). All the d$I$/d$V$ data, regardless of the direction of magnetic field, exhibit similar $V$-shape and field dependence, indicating the same physical origin and robust intrinsic character. Interestingly, the behavior discovered in CaRh$_2$As$_2$ is similar to that reported for pressurized underdoped CeCoIn$_5$, where it was attributed to the emergence of a new-type pseudogap[18].

Fig. 3 displays the Fano normalized d$I$/d$V$ data of CeRh$_2$As$_2$ (#2) in a form of color contour plots of temperature or magnetic field ($B$//$c$) versus bias voltage. In a fixed field of 0, 5 and 7 T, Figs. 3(a)-(c), a feather-shaped zone is located near the zero-bias voltage, $T_g$ decreases to ~ 6 K at 5 T and ~ 3 K at 7 T, and the corresponding bias voltages decay quickly. These results demonstrate a monotonic suppression of the pseudogap by magnetic field. In the center of the feather zone, the color gradient suggests that the pseudogap is filled up smoothly and almost independently of temperature, similar to cuprates[19] and CeCoIn$_5$[18,31,32]. In contrast, at a fixed temperature of 0.1, 0.5, and 1.7 K, Figs. 3(d)-(f), there is a significant enhancement of signal in the central area having a cylinder-like shape. With increasing temperature, $B_g$ slightly decreases and $2\Delta_g$ shows a significant enlargement from ~ 0.95 meV at 0.1 K to ~ 3 meV at 1.7 K. These characteristics are consistent with broadening of the $V$-shaped peak, as discussed above.

Based on the collected PCS data, the magnetic field-temperature dependencies of the superconducting state and the pseudogap were integrated into the $B$-$T$ phase diagrams of CeRh$_2$As$_2$, constructed for $B//c$, Fig. 4(a), and $B \perp c$, Fig. 4(b), configurations. The colored regions represent the dip depth of the $dI/dV$ spectra. As seen clearly in Fig. 4(a), the critical temperature $T_c^{zero}$ of three samples (S1#, S2#, S3#) shows almost identical variations with field, fully consistent with the literature data derived from $C_p(T)$, $\rho(T)$ and $\chi(T)$. The $T_c^{ZBCP}$ boundary, determined from the ZBCP($T$), overlaps with $T_c^{zero}$ for $B < 4$ T (SC1 phase), but decreases faster and deviates significantly from the $T_c^{zero}$ line in stronger fields (SC2 phase). In the latter region, $T_c^{ZBCP}$ quickly decreases and becomes undetectable above 7 T, which is only half of $B_{c2} \approx 14$ T of the SC2 phase. This finding suggests that the superconducting signal in the $dI/dV$ spectra seen in the high-field region should rather be considered an extension of the SC1 phase and is probably not related to the presence of the SC2 phase. If so, the SC2 phase showed no features in the $dI/dV$ spectra, which might be attributed to its odd-parity superconducting symmetry. To validate such hypothesis, further in-depth experimental and theoretical studies are required. In the case of $B \perp c$, a perfect overlap between the $T_c^{ZBCP}$ and $T_c^{zero}$ lines determined for two CeRh$_2$As$_2$ samples (S1#, S2#) was found, Fig. 4(b), and the estimated value $B_{c2} \approx 2.0$T at $T = 50$mK is very similar to that obtained from the bulk thermodynamic and transport measurements. This result further corroborates the PCS technique as an appropriate experimental technique for characterization of the SC1 phase of CeRh$_2$As$_2$.

Above $T_c$, the pseudogap features emerge and are gradually suppressed with increasing temperatures and magnetic fields. For both configurations of the magnetic field relative to the crystallographic axes the corresponding phase boundary in the $B - T$ diagrams are similar and yield the limiting values $T_g \approx$ 8-9 K and $B_g \approx 9.0 \pm 0.5$ T. Notably, the $dI/dV$ spectra, while being highly sensitive to temperature and external field, are sample independent, in contrast to Fano effect signals. Another important observation is an entirely different behavior of $T_g(B)$ measured for $B \perp c$ with respect to the evolution of $T_0(B)$ that increases monotonically with increasing field, especially in high field region, Fig. 4(b). It implies that the $V$-shaped feature in the $dI/dV$ spectra is not related to the $T_0$ phase, and thus supports the pseudogap scenario.

Our discovery of the pseudogap in $CeRh_2As_2$ provides an excellent new material platform to elucidate its poorly understood microscopic mechanism and the relationship between this phenomenon and unconventional superconductivity in strongly correlated electron systems. So far, the emergence of pseudogap has been observed in high-$T_c$ cuprates[33] with appreciable energy gaps persisting in a wide temperature range above $T_c$. The interplay between pseudogap and high-$T_c$ SC has been a subject of extensive investigations over the past decades. Now, it is generally believed that the pseudogap can originate from two mechanisms[34], i.e., (i) the electron pre-pairing above $T_c$ in the Bardeen-Cooper-Schrieffer (BCS) to Bose-Einstein condensation (BEC) crossover picture, or (ii) the presence of various competing quantum electronic orders, such as the striped AFM order or paired density wave (PDW) order. However, due to the complexity of high-$T_c$ cuprates and possible intertwining of these two mechanisms, the underlying origin of the pseudogap remains elusive. In addition to the unconventional SCs, some recent studies proposed new candidates for pseudogap phenomena, including uniform Fermi gases with strong interactions[35] and magnetic transition materials. This makes it even more complex to unveil the principal mechanism of the pseudogap formation.

In an attempt to identify a possible mechanism for the opening of pseudogap in $CeRh_2As_2$, it seems worthwhile to compare its bulk physical properties with those of other heavy fermion systems. At first, the presence of a hidden order (HO) as seen in $URu_2Si_2$[36] can be excluded because there are no obvious anomalies in the physical properties of $C_p(T)$, $\rho(T)$ and $\chi(T)$ around $T_g$. In the absence of obvious phase transition, the pseudogap can be caused by Kondo hybridization as proposed in $CeCoIn_5$[18]. However, the observed strong variations of pseudogap with temperature and magnetic field don't comply with the latter scenario. Instead, previous studies have suggested that $CeRh_2As_2$ approaches quantum criticality due to spin density wave (SDW) or QDW orders, associated with enhanced spin fluctuations and divergent effective electron mass[21,22]. Thus, quantum fluctuations might serve as an important trigger for strongly coupled superconducting pairing (larger $2\Delta_{SC}/k_BT_c$) in this material, as well as for the emerging pseudogap in the normal state ($2\Delta_g \approx$ 0.95-3.0 meV), reminiscent of high-$T_c$ cuprates. It must be stressed again that further experimental and theoretical investigations on its microscopic electronic structures are essential to clarify the underlying mechanisms of the pseudogap opening in $CeRh_2As_2$.

In summary, by employing soft PCS technique we discovered the emergence of pseudogap features in the putative multiphase HFSC $CeRh_2As_2$. The pseudogap spans a wide range of $2\Delta_g \approx$ 0.95-3.0 meV and persists up temperatures as high $T_g \approx$ 8-9 K. Its magnitude diminishes upon application of external magnetic field but remains finite up to a critical field $B_g \approx 9.0\pm 0.5$ T. From the PCS spectra we determined the width of the superconducting energy gap $2\Delta_{SC} \approx 0.24$ meV at 50 mK. This value implies an extremely strong superconducting pairing strength $2\Delta_{SC}/k_BT_c \approx 8.8$, which is comparable to that of cuprates and iron-based high-$T_c$ SCs. Our findings indicate that CeRh$_2$As$_2$ can be considered as a novel material platform to study the interplay of unconventional SC and pseudogap phenomena.

## Acknowledgements

We thank Dr. Tian Le, Xing-yuan Hou, Hai Zi and Feng-rui Shi for fruitful discussions. This work is supported by the National Key Research and Development Program of China (Grants No. 2024YFA1408400, No. 2023YFA1406100, No. 2023YFA1607400, No. 2022YFA1403800, No. 2022YFA1403203), the National Natural Science Foundation of China (Grants No. 12474055, No. 12404067, No. 12025408, No. U23A6003), the Strategic Priority Research Program of CAS (Grant No. XDB33000000), the CAS President's International Fellowship Initiative (Grants No. 2024PG0003), and the Outstanding member of Youth Promotion Association of Chinese Academy of Sciences (Grants No. Y2022004). This work was supported by the CAC station of Synergetic Extreme Condition User Facility (SECUF，https://cstr.cn/31123.02.SECUF).

# Figure Captions

**FIG. 1.** Soft PCS measurements of CeRh$_2$As$_2$ (S1#). (a) Quasi-four-probe configuration of the electrical leads. (b) The d$I$/d$V$ spectra taken with the AC current $I_{AC}$ = 50 μA. (c) The normalized (d$I$/d$V$)/(d$I$/d$V$)$_{0.3K}$ spectra. (d) and (f) The Andreev reflection spectra measured at $T$ = 50 mK in various magnetic fields applied along and perpendicular to the crystallographic $c$-axis, respectively. (e) and (g) The temperature dependencies of the zero-bias conductance peak measured in various fields $B//c$ and $B \perp c$, respectively.

**FIG. 2.** PCS measurements of CeRh$_2$As$_2$ (S1#, S2#) (a) The d$I$/d$V$ spectra collected in the temperature range 60 mK-12 K with the current $I_{AC}$ = 100 μA. For clarity, the individual spectra were shifted up equidistantly. The grey solid lines represent the fits using the Fano resonance model (see the main text and SM). (b) The normalized (d$I$/d$V$)/(d$I$/d$V$)$_{Fano}$ spectra obtained by subtracting the Fano contributions. The arrows near the zero-bias voltage mark the pseudogap width. (c) and (d) The d$I$/d$V$ spectra measured with $B // c$ and $I_{AC}$ = 100 μA at $T$ = 0.1 K and 0.5 K, respectively. (e) The d$I$/d$V$ spectra measured at $T$ = 0.5 K with $B \perp c$ and $I_{AC}$ = 100 μA.

**FIG. 3.** The normalized (d$I$/d$V$)/(d$I$/d$V$)$_{Fano}$ spectra of CeRh$_2$As$_2$ (S1#, S2#) in a form of the color contour plot with the coordinates temperature or magnetic field ($B//c$) versus bias voltage. (a), (b) and (c) The data taken in fixed fields $B$ = 0, 5, and 7 T, respectively. (d), (e) and (f) The data taken at fixed temperatures $T$ = 0.1, 0.5 and 1.7 K, respectively. The dashed lines mark the boundaries.

**FIG. 4.** Temperature-magnetic field phase diagrams of CeRh$_2$As$_2$ (S1#, S2#, S3#) constructed for (a) $B//c$ and (b) $B \perp c$. The values of $T_c^{zero}$ and $T_c^{ZBCP}$ were derived from the $\rho(T)$ and ZBCP($T$) data, respectively. The $T_c$ and $T_0$ values were adopted from $\chi(T)$ and $C_p(T)$ reported in Ref. [1]. The $T_0$ boundary was plotted based on the data shown in Ref. [3,5,8,20].

# Figures
# Figure 1

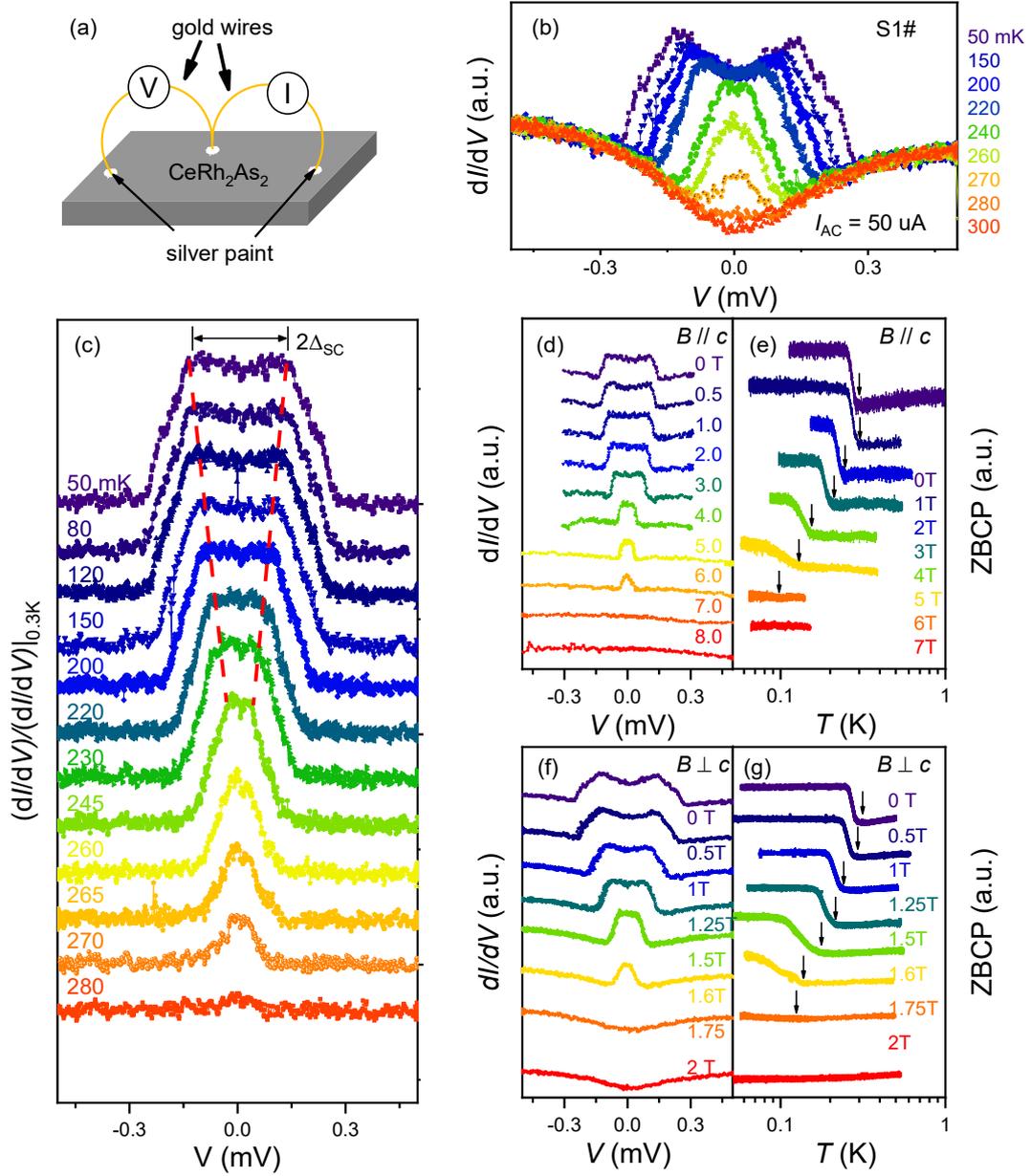

# Figure 2

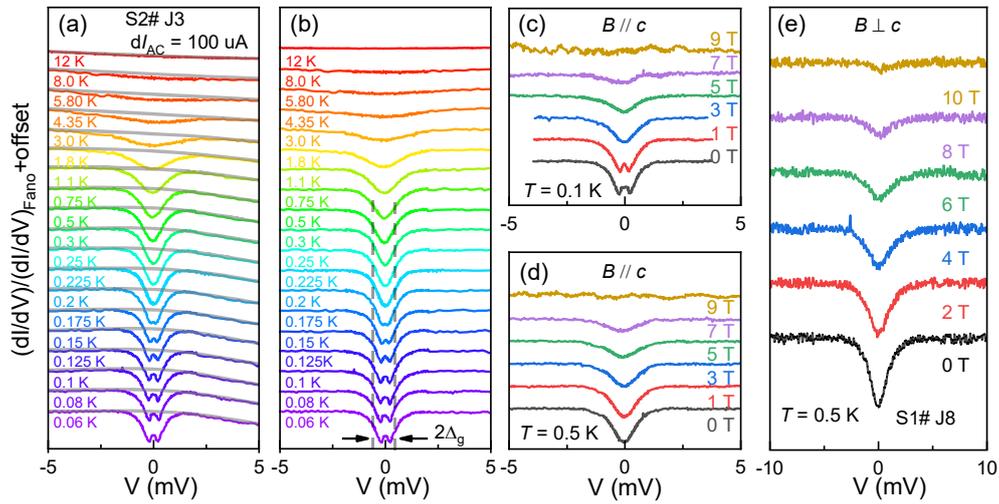

# Figure 3

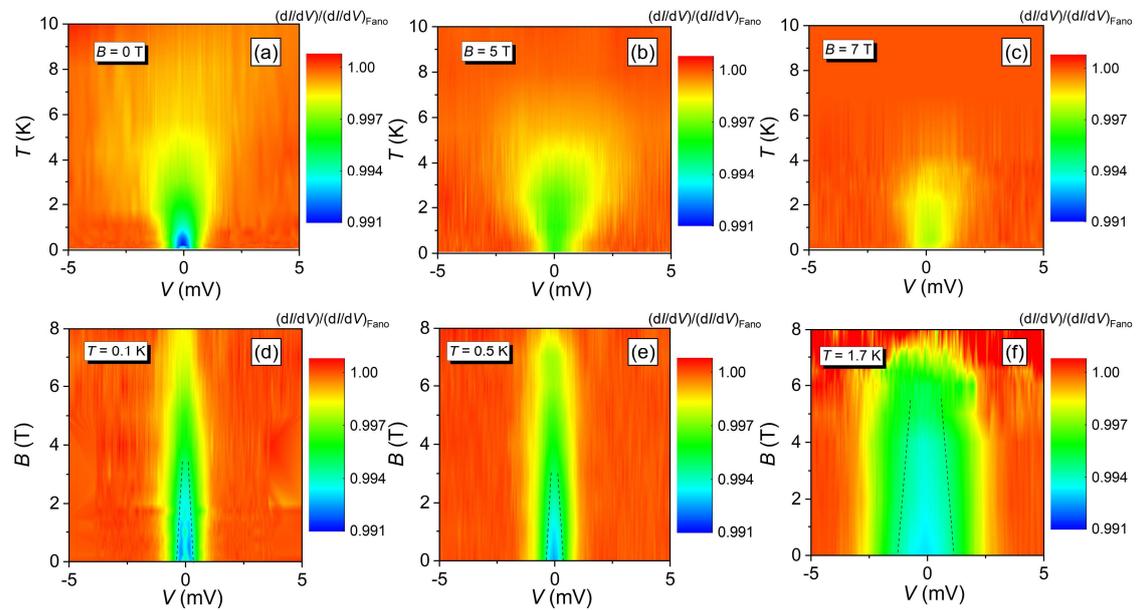

**Figure 4**

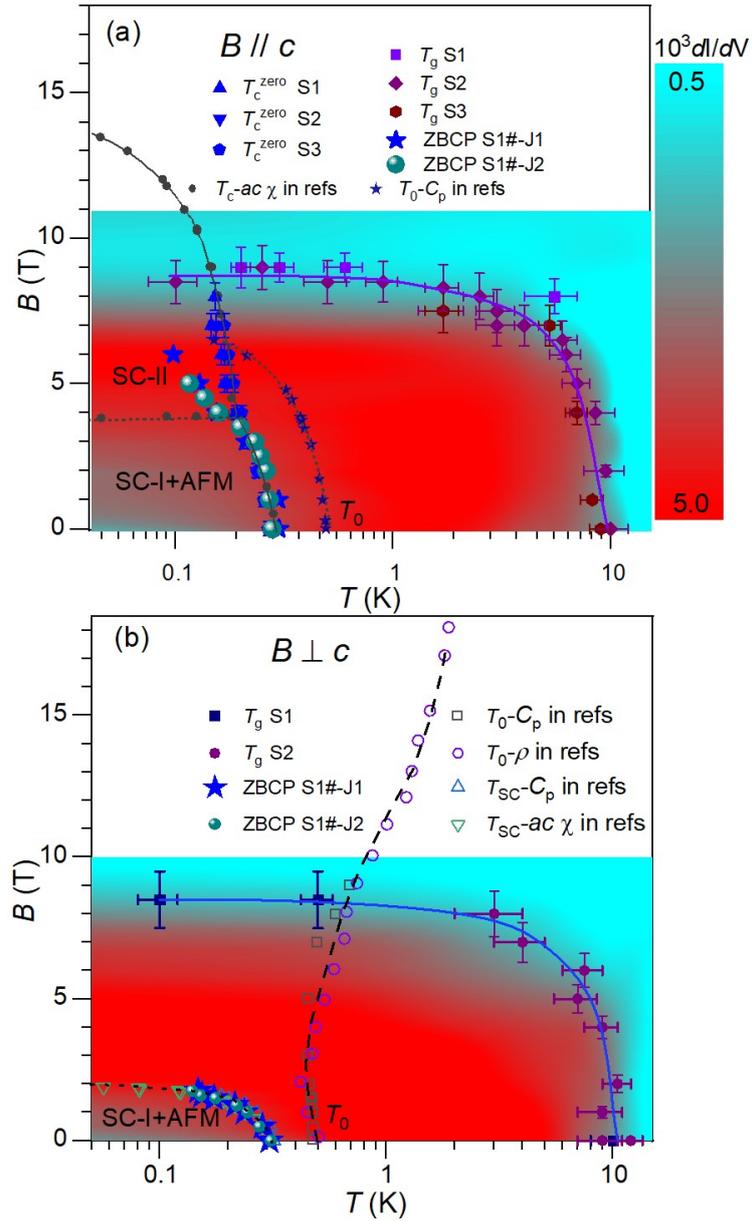